\documentclass[12pt]{article}
\usepackage{amssymb}
\newif\ifusepdf
\usepdffalse  
\ifusepdf
\newif\ifpdf
\ifx\pdfoutput\undefined
   \pdffalse
   \usepackage{cite}
 \else
   \pdfoutput=1
   \pdftrue
  \usepackage[pdftex]{hyperref}
   \providecommand{\href}[2]{#2}
\newcommand\email[1]{{\tt\href{mailto:#1}{\textbf{\texttt{#1}}}}}
\fi
\else
\newcommand\email[1]{\textbf{\texttt{#1}}}
\newcommand\href[2]{#2}
\fi
\newcommand{\sign}{\mathop{\rm sign}}
\newcommand{\SU}{\mathop{\rm SU}}
\newcommand{\SO}{\mathop{\rm SO}}
\newcommand{\U}{\mathop{\rm U}}
\newcommand{\Sp}{\mathop{\rm Sp}}

\newcommand{\OSp}{\mathop{\rm OSp}}

\newcommand{\Spin}{\mathop{\rm {}Spin}}

\newcommand{\be}{\begin{equation}}

\newcommand{\ee}{\end{equation}}
\newcommand{\h}{\eta}
\newcommand{\bea}{\begin{eqnarray}}
\newcommand{\eea}{\end{eqnarray}}

\newcommand{\nn}{\nonumber}

 \def\slash#1{\setbox0=\hbox{$#1$}#1\hskip-\wd0\dimen0=5pt\advance
       \dimen0 by-\ht0\advance\dimen0 by\dp0\lower0.5\dimen0\hbox
         to\wd0{\hss\sl/\/\hss}}

\newcommand{\m}{\mu}
\newcommand{\n}{\nu}

\newcommand{\bi}{\begin{enumerate}}
\newcommand{\ei}{\end{enumerate}}
\newcommand{\A}{\alpha} \newcommand{\B}{\beta} 
  
\newcommand{\ep}{\epsilon}

\newcommand{\lam}{\lambda}
      \newcommand{\s}{\sigma}

\def\6{\partial}
\def\7{\tilde}
\def\8{\hat}

\def\CC{{\cal C}}

\def\t{\tilde}

\def\={{\;=\;}}\def\+{{\;+\;}}

\newcommand{\slP}{/ {\hskip-0.25cm{P}}}

\renewcommand{\varepsilon }{\epsilon }
\begin{document}
\begin{titlepage}
\begin{flushright}
ICCUB-08-135, KEK-1285, KUL-TF-08/32, UB-ECM-PF-08/19, Toho-CP-0889\\
arXiv:0812.1982 [hep-th]
\end{flushright}
\vspace{.5cm}
\begin{center}
\baselineskip=16pt {\LARGE  \bf{Vector Supersymmetry: Casimir operators\\
\vskip 0.2cm and contraction from $\OSp(3,2\,|2)$ }}\\
\vskip 5mm

{\large Roberto Casalbuoni$^1$, Federico Elmetti$^2$,
Joaquim Gomis$^3$, \\[2mm]
Kiyoshi Kamimura$^4$, Laura Tamassia$^2$
 and Antoine Van Proeyen$^2$}
\vskip 5mm

{\small {\it $^1$ Department of Physics, University of Florence,\\
         INFN-Florence and Galileo Galilei Institute,
          Florence, Italy} \\
          \email{casalbuoni@fi.infn.it}\\\vspace{4mm}
$^2$ {\it Instituut voor Theoretische Fysica, Katholieke Universiteit Leuven,\\
Celestijnenlaan 200D B-3001 Leuven, Belgium} \\
\email{federico.elmetti@fys.kuleuven.be},
\email{laura.tamassia@fys.kuleuven.be}, \email{antoine.vanproeyen@fys.kuleuven.be}\\\vspace{4mm}
$^3$ {\it Departament d'Estructura i Constituents de la Mat{\`e}ria and
Institut de Ci{\`e}ncies del Cosmos, Universitat de
Barcelona, Diagonal 647, 08028 Barcelona, Spain}\\\vspace{1mm}
{\it High Energy Accelerator Research Organization (KEK),\\ Tsukuba,
Ibaraki, 305-0801 Japan}\\ \email{gomis@ecm.ub.es}\\
\vspace{4mm}
$^4$ {\it Department of Physics, Toho University, Funabashi, 274-8510
Japan}}\\ \email{kamimura@ph.sci.toho-u.ac.jp}
\end{center}
\vskip 3mm
\begin{center}
{\bf Abstract}
\end{center}
{\small We study some algebraic properties of the `vector supersymmetry'
(VSUSY) algebra, a graded extension of the four-dimensional Poincar{\'e}
algebra with two odd generators, a vector and a scalar, and two central
charges. The anticommutator between the two odd generators gives the
four-momentum operator, from which the name vector supersymmetry. We
construct the Casimir operators for this algebra and we show how both
algebra and Casimirs can be derived by contraction from the simple
orthosymplectic algebra $\OSp(3,2\,|2)$. In particular, we construct the
analogue of superspin for vector supersymmetry and we show that, due to
the algebraic structure of the Casimirs, the multiplets are either
doublets of spin $(s,s+1)$ or two spin 1/2 states. Finally, we identify
an odd operator, which is an invariant in a subclass of representations
where a BPS-like algebraic relation between the mass and the values of
the central charges is satisfied.}
\end{titlepage}
\addtocounter{page}{1}
 \tableofcontents{}
\newpage


\section{Introduction\label{sec:0}}

The vector supersymmetry (VSUSY) algebra is a graded extension of the
Poincar{\'e} algebra in four dimensions. Two fermionic operators are added,
an odd Lorentz vector and an odd Lorentz scalar, and two central charges
are allowed. The anticommutator between vector and scalar odd generators
gives the four-momentum vector, from which the name vector supersymmetry.

This algebra was first introduced in \cite{Barducci:1976qu} in 1976, with
the purpose of obtaining a pseudoclassical description of the Dirac
equation. However, to our knowledge its general algebraic properties have
never been studied in detail. Due to the prominent role of supersymmetry
algebras in field and string theories, VSUSY representations and possible
realizations in physical models are worth exploring. In any case, it
would be interesting to compare this alternative to ordinary
supersymmetry to understand what the essential ingredients in
supersymmetry are.

The main difference between vector and ordinary supersymmetry is that the
odd generators of VSUSY do not satisfy the spin-statistics rule. This is
not necessarily a problem for the construction of physical models with
underlying VSUSY. In fact, a first example is the model for the spinning
particle constructed in \cite{Casalbuoni:2008iy}, where a quantization
procedure preserving the underlying VSUSY has been applied, while the
authors in \cite{Barducci:1976qu} had to break the symmetry.\footnote{See
also \cite{Gamboa:1997fs}.}

There is also an interesting connection between VSUSY and topological
field theories. In fact, an Euclidean version of VSUSY appears as a
subalgebra of the symmetry algebra underlying topological $\mathcal{N}=2$
Yang-Mills theories. Supersymmetry with odd vector generators was studied
after Witten \cite{Witten:1988ze}, who, in 1988, introduced topological
$\mathcal{N}=2$ Yang-Mills theories by performing a topological twist.
After this twist, the fermionic generators become a vector, a scalar and
an anti-selfdual tensor \cite{Alvarez:1994ii,Kato:2005fj}. After
truncation of the anti-selfdual sector, the twisted algebra coincides
with the Euclidean VSUSY algebra, in the special case when the two
central charges of VSUSY are identified. Twisted topological algebras
have proven to be useful in the study of renormalization properties of
topological field theories \cite{Birmingham:1988bx,Delduc:1989ft}.
Moreover, a superspace formalism has been developed for these topological
theories, see for example
\cite{Alvarez:1994ii,Kato:2005fj,Baulieu:2008at} and references therein.

An understanding of the physical content of theories with underlying
VSUSY can be achieved by classifying VSUSY representations. A first step
in this direction is to identify the Casimir operators of the algebra. We
find that there are four Casimirs, $P ^2$, $Z$, $\tilde Z$ and $\hat W^2$.
Here $P_\mu$ is the four-momentum, $Z$ and $\tilde Z$ are the central
charges and $\hat W^2$ is the square of $\hat W_\mu $, the analogue of
the superspin vector of ordinary supersymmetry \cite{Salam:1974yz}, which
is a generalization of the Pauli-Lubanski vector. The Pauli-Lubanski
vector, which determines the spin of particles, can be written as a sum
of this vector $\hat W_\mu$ and another one, $W_{C\,\mu}$. The latter is
constructed in terms of the generators of the supertranslation subalgebra
of VSUSY. It squares to $P^2$ and has spin $\frac12$. As a result of this
structure, a VSUSY irreducible representation contains two particles of
Lorentz spin $s=|Y\pm\frac12|$, where $Y$ is the superspin having integer
or half integer value. In particular, in the case $Y=0$, one has two spin
1/2 states. The described structure holds for the generic case with
nonvanishing central charges, on which we focus in this paper. In the
case of vanishing central charges the contribution of the Pauli-Lubanski
vector vanishes in the expression of the superspin. We leave this special
case for future work.

A comparison with the case of ordinary supersymmetry is in order. We
observe that also in that case the Pauli-Lubanski vector can be written
as the sum of the superspin vector and another spin vector (see for
example \cite{Pasqua:2004vq}). The main difference between the two cases
is that for ordinary supersymmetry the second spin vector does not square
to a Casimir.

In case the mass and central charges satisfy a BPS-like relation, we
identify an odd operator that behaves as a Casimir. In general, odd
Casimirs can be present when the odd generators are scalars or vectors.
In the case of VSUSY, there is no odd Casimir but we have constructed an
odd nilpotent operator, invariant in the subclass of representations
where the mentioned relation between mass and central charges is
satisfied.

A good strategy to gain some understanding about a new algebra is to try
to relate it to some other, better-known algebra. In this direction, we
show that VSUSY arises as a contraction of the superalgebra
$\OSp(3,2\,|2)$. $\OSp$ algebras are very simple generalizations of $\SO$
or $\Sp$ algebras. $\SO$ algebras have a symmetric metric, $\Sp$ algebras
have an antisymmetric metric and $\OSp$ algebras have a `graded
symmetric' metric. $\OSp$ algebras are natural candidates for an
embedding of VSUSY. The reason is that we need our VSUSY fermions to
appear as vectors (or scalars) of the Lorentz group. In general, for
$\OSp(M|N)$ algebras the fermions are vectors of $\SO(N)$ and of
$\Sp(M)$. We also need a bosonic factor to embed the two central charges.
Therefore, we will use the embedding in $\OSp(3,2\,|2)$, whose bosonic
part is $\SO(3,2)\times \Sp(2)$. The latter factor  can host the central
charges we want to include.

We have also rederived the VSUSY Casimirs by contraction from
$\OSp(3,2\,|2)$. This procedure turns out to be rather nontrivial, mainly
due to the fact that VSUSY has two central charges. In fact,
$\OSp(3,2\,|2)$ has three independent Casimirs, while, as mentioned
above, VSUSY has four. Therefore, some nontrivial combination of the
$\OSp$ Casimirs together with a careful limit procedure have to be
performed to derive the VSUSY superspin.

It is interesting to compare this result to the case of ordinary
$\mathcal{N}=1$ SUSY in four dimensions. The corresponding superalgebra
can be derived by contraction from $\OSp(1|4)$ \cite{Hlavaty:1980md},
which has two independent Casimirs. Contraction of the first leads to
$P^2$, while contraction of the second leads in fact to the superspin
operator, but in a form that is not at all familiar to physicists (see
for instance \cite{Gates:1983nr,Buchbinder:1998qv,paul}). Therefore,
in the case of ordinary $\mathcal{N}=1$ SUSY, all Casimir operators can
be obtained by direct contraction from the $\OSp$ embedding algebra.

The paper is organized as follows. In Section \ref{ss:VSUSYandCasimir} we
introduce the VSUSY algebra and we present its Casimir operators.
Moreover, we compare our result to the analogue for ordinary
supersymmetry. In Section \ref{ss:VSUSYOSp} we briefly introduce the
$\OSp(3,2\,|2)$ algebra and we show how to derive VSUSY by a contraction
procedure. In Section \ref{ss:contractionCasimirs} we discuss the
contraction of the Casimir operators and specially how to derive the
analogue of superspin for VSUSY. In Section \ref{ss:conclusions}, we
summarize our results and we present our plans for future work. In
Appendix \ref{app:CasimirsVSUSY}, we show in detail how to derive all
independent Casimirs of VSUSY. In Appendix \ref{app:OSp}, a brief
technical introduction to $\OSp$ algebras is given and our conventions
are stated. Finally, in Appendix \ref{app:OSpcontraction}, some
commutation relations useful for  the contraction procedure are given.

\section{VSUSY and its Casimir operators}
 \label{ss:VSUSYandCasimir}
\subsection{VSUSY algebra} \label{ss:VSUSYalg}

The vector supersymmetry (VSUSY) algebra in 4 dimensions is a graded
algebra defined by the even generators $P_\mu$, $M_{\mu\nu}$, $Z$,
$\tilde Z$, and by the odd generators $G_\mu$ and $G_5$. The algebra of
the even generators $P_\mu$ and $M_{\mu\nu}$ is the usual Poincar{\'e}
algebra, whereas the odd generators behave respectively as a four-vector
and a scalar under the Lorentz group and are translationally invariant.
The non-zero (anti)commutation relations are  
\begin{eqnarray}
&&[M_{\mu\nu},M_{\rho\sigma}]_-=\eta_{\nu\rho}M_{\mu\sigma}+
\eta_{\mu\sigma}M_{\nu\rho}-\eta_{\nu\sigma}M_{\mu\rho}-
\eta_{\mu\rho}M_{\nu\sigma}\,,\nonumber\\
&&[M_{\mu\nu},P_\rho]_-=-\eta_{\mu\rho}P_\nu+\eta_{\nu\rho}P_\mu\,,\qquad
[M_{\mu\nu},G_\rho]_-=-\eta_{\mu\rho}G_\nu+\eta_{\nu\rho}G_\mu\,,\nonumber\\
&&[G_\mu,G_\nu ]_+=\eta_{\mu\nu}Z\,,\qquad [G_5,G_5 ]_+=\tilde Z
\,,\qquad [G_\mu,G_5]_+=-P_\mu\,,
 \label{vsusyalgebra}
\end{eqnarray}
where we use $[\cdot ,\cdot ]_-$ for commutators and $[\cdot ,\cdot ]_+$
for anticommutators. Here and after, the following conventions will be
used for the metric and the Levi Civita tensor
\begin{equation}\eta_{\mu\nu}={\rm diag}(-1,+1,+1,+1)\,,\qquad
\ep^{0123}=-\ep_{0123}=1\,. \label{conventions1s5}
\end{equation}
Two remarks should be made concerning these equations. First, the VSUSY
algebra makes perfect sense as a real algebra. The `$i$' factors
appearing in \cite{Casalbuoni:2008iy} were introduced for physical
reasons. Since in this paper we are mainly concerned with algebraic
properties, we omit them. The complex version of the commutation
relations can be obtained by replacing $M_{\mu\nu}$ with $iM_{\mu \nu }$
in (\ref{vsusyalgebra}). Second, $Z$ and $\tilde Z$ are central charges.
Therefore, in any representation they can be considered as numbers.

In the case of non-vanishing $Z$ and $\tilde Z$, only their signs and
their product are relevant. This can be seen by rescaling
\begin{equation}
\widehat{G}_\mu=\frac 1\alpha G_\mu\,,\qquad \widehat{G}_5=\alpha G_5\,.
\label{scalingalpha}
\end{equation}
As a result, the odd sector of the algebra becomes
\begin{equation}
[\widehat{G}_\mu,\widehat{G}_\nu]_+=\eta_{\mu\nu}\widehat Z=\eta_{\mu\nu}\frac{1}{\alpha ^2} Z\,,\qquad
[\widehat{G}_5,\widehat{G}_5]_+=\widehat{\tilde Z}=\alpha ^2\tilde Z\,,\qquad
[\widehat{G}_\mu,\widehat{G}_5]_+=-P_\mu\,.
\end{equation}
We can choose $\alpha^2=\sqrt{\frac{|Z|}{|\tilde Z|}}$, so that the two
new central charges have the same absolute value $|\widehat
Z|=|\widehat{\tilde Z}|=\sqrt{|Z\tilde Z|}=c$ and the algebra is
specified by the value of $c$ and by the signs of $Z$ and $\tilde Z$ as
\begin{equation}
[\widehat{G}_\mu,\widehat{G}_\nu]_+=\eta_{\mu\nu}{\rm
sign}(Z)\,c\,,\qquad [\widehat{G}_5,\widehat{G}_5]_+={\rm
sign}(\tilde Z)\,c\,,\qquad
[\widehat{G}_\mu,\widehat{G}_5]_+=-P_\mu\,.
\end{equation}
\subsection{Casimir operators}

The central charges $Z$ and $\tilde{Z}$ are trivial Casimirs of VSUSY. It
is also easy to see that $P^2$ is a Casimir for VSUSY. As in the case of
ordinary supersymmetry, we expect that an analogue of superspin
\cite{Salam:1974yz} could be constructed by starting from a
generalization of the Pauli-Lubanski vector
\begin{equation}
W^\mu=\frac{1}{2}\epsilon^{\mu\nu\rho\sigma}P_\nu M_{\rho\sigma}\,.
\label{PLW}
\end{equation}
$W^2$ itself is not a Casimir. As a result, particles of different
Lorentz spin will appear in the same multiplet. The correct VSUSY
generalization of the Pauli-Lubanski vector is
\begin{equation}
\hat W^\mu=\frac{1}{2}
\epsilon^{\mu\nu\rho\sigma}P_\nu (ZM_{\rho\sigma}-G_{\rho} G_{\sigma})\,,
\label{defhW}
\end{equation}
whose square $\hat W^2$ is a Casimir. More details concerning how to
derive these Casimir operators and how to prove that there are no further
independent ones are given in Appendix \ref{app:CasimirsVSUSY}.

By introducing the new vector
\begin{equation}
W_C^\mu=\frac{1}{2}\ep^{\mu\nu\rho\s}P_\nu G_{\rho} G_{\s}\,,
\label{PRCC}
\end{equation}
one can rewrite formula (\ref{defhW}) as
\begin{equation}
Z W^\mu=\hat W^\mu+W_C^\mu\,.\label{spinsum}
\end{equation}
One can easily prove that $W^2_C$ is also a Casimir. However, it is not
independent, since
\begin{equation}
 W_C^2=Z^2 P^2\;\frac34\,.
\label{C-spin}
\end{equation}

From now on, we are implicitly considering the case of representations
with $Z\neq 0$. In that case $Z$ is just a number and can be divided out.
The three vectors $W_*^\mu=\frac{\hat W^\mu}{Z}$, $W^\mu$,
$\frac{W_C^\mu}{Z}$ all commute with $P_\mu $ and $G_5$ and verify the
relation
\begin{equation}
\left[W_*^\mu,W_*^\nu\right]_-=\epsilon^{\mu\nu\rho\sigma}P_\rho
W_{*\sigma}\,.\label{spinalgA}
\end{equation}
Therefore, in the rest frame of the massive states where $P^2=-m^2$, they
satisfy the rotation algebra
\begin{eqnarray}
\left[ \frac{W_*^i}{m},\frac{W_*^j}{m}
\right]_-=\ep^{ijk}\,\frac{W_{*k}}{m}\,,
\end{eqnarray}
and define three different spins. The superspin $Y$ labels the
eigenvalues $-m^2Z^2Y(Y+1)$ of the Casimir ${\hat W}^2$. The spin
associated to $W_C^2$ (C-spin) is fixed to $1/2$, as one can see from
(\ref{C-spin}). Finally, we denote the usual Lorentz spin by $s$.

On the other hand, only $W_*^\mu=\frac{1}{Z}\hat W^\mu$ commutes with
$G_\lam$, and thus only the superspin $Y$ characterizes a multiplet.
Since
\be
 \left[{\hat W}^\mu,W_C^\nu\right]_-=0\,,\label{indepspin}
 \ee
one can immediately obtain the particle content of a VSUSY multiplet by
using the formal theory of addition of angular momenta applied to
(\ref{spinsum}). As a result, a multiplet of superspin $Y$ contains two
particles of Lorentz spin $Y\pm 1/2$, for $Y>0$ integer or half-integer.
In the degenerate case of superspin $Y=0$, the multiplet consists of two
spin $1/2$ states. In particular, we observe that a VSUSY multiplet
contains either only particles of half-integer Lorentz spin or only
particles of integer Lorentz spin. The spinning particle constructed in
\cite{Casalbuoni:2008iy} is a realization of the degenerate case $Y=0$.

To summarize, we draw the following table. The last column refers to the
particle model in \cite{Casalbuoni:2008iy}. We stress that the eigenvalues
appearing in the table are all negative due to the fact that we have chosen
a real algebra and antihermitian operators.
\vskip 3mm
\begin{tabular}{|c|c|c|c|c|}
  \hline
 & eigenvalue &  &  & vector superparticle  \\
  \hline
  $\frac{1}{Z^2}\8W^2$ & $-m^2\;Y(Y+1)$ & $\mathrm{superspin}=Y$& $\mathrm{Casimir}$& $Y=0 $ \\
 &  & & &  \\
  $\frac{1}{Z^2} W_C^2$ & $-m^2\;\frac34$ & $\rm{C\,\,spin}=\frac12 $ &$\mathrm{Casimir}$&  $ C=\frac12 $ \\
   &  & & &  \\
$W^2$& $-m^2\;s(s+1)$ & $\mathrm{Lorentz\; spin}=s=|Y\pm\frac12|$ &
$\mathrm{not\;Casimir}$ & $  s=\frac12 $ \\
  \hline
\end{tabular}

\subsection{Superspin and Lorentz spin for ordinary supersymmetry}

Both for ordinary supersymmetry and VSUSY it is possible to construct a
superspin Casimir operator starting from a generalization of the
Pauli-Lubanski vector. In this section we would like to revisit the
construction of the superspin Casimir for ordinary supersymmetry along
the lines of what we have done for VSUSY.

We normalize the supersymmetries by assuming that the anticommutator between
components $Q^\alpha $ of the spinorial supersymmetry charge $Q$ has the form
\begin{equation}
  \left[ Q^\alpha ,Q^\beta \right] = 2 \left( \gamma ^\mu C^{-1}\right)^{\alpha \beta
  }\,,
 \label{QQanticomm}
\end{equation}
where $C$ is the antisymmetric charge conjugation matrix used to define
$\bar Q =Q^T C$. The suitable generalization of the Pauli-Lubanski vector
$W^\mu $ reads \cite{Salam:1974yz}
\begin{equation}
  Z^\mu = W^\mu -\frac{1}{8}\bar Q\gamma^\mu\gamma_5 Q\,,
 \label{defZmususy}
\end{equation}
where $\gamma_5=\gamma_0 \gamma_1 \gamma_2 \gamma_3$, so that $(\gamma
_5)^2=-1$ and $\gamma ^{\mu \nu }\gamma _5= -\frac{1}{2} \varepsilon
^{\mu \nu \rho \sigma }\gamma _{\rho \sigma}$.
Its commutator with the supersymmetry charge is
\begin{equation}
  [Z^\mu ,Q]_-= \frac{1}{2} P^\mu Q \,,
 \label{ZQproptoP}
\end{equation}
such that $Z^\mu P^\nu- Z^\nu P^\mu $ commutes with $Q$. The superspin
Casimir operator is usually written in the literature in the form
\be
\CC=\frac12\left( Z_\mu  P_\nu- Z_\nu  P_\mu\right) \left( Z^\mu  P^\nu
-Z^\nu P^\mu\right) =Z^2 P^2-(Z\cdot P)^2\,. \label{CCsusy}
\ee
In the representations
where the Casimir $P^2$ is nonvanishing, we can equivalently consider
$\frac{\CC}{P^2}$ as the superspin Casimir. The latter can be expressed
as the square of the vector
\be
\hat W^\mu=Z^\nu\left({\delta_\nu}^\mu-\frac{P_\nu P^\mu}{P^2}\right)=
W^\mu-\frac{1}{8}\bar Q\gamma^\nu\gamma_5
Q\left({\delta_\nu}^\mu-\frac{P_\nu P^\mu}{P^2}\right)\,.
\ee
Then, exactly as in the case of VSUSY, the Pauli-Lubanski vector is the sum of
two commuting vectors \cite{Pasqua:2004vq},
\be
W^\mu=\hat W^\mu+\tilde{W}_C^\mu\,. \label{spindec}
\ee
 where
\begin{equation}
  \tilde{W}_C^\mu=\frac{1}{8}\bar Q\gamma^\nu\gamma_5
Q\left({\delta_\nu}^\mu-\frac{P_\nu P^\mu}{P^2}\right)\,.
\end{equation}
Moreover, as in the VSUSY case, the three vectors $W_*^\mu= W^\mu,\hat
W^\mu, \tilde{W}_C^\mu$ satisfy
\be
\left[
W_*^\mu,W_*^\nu\right]_-=\ep^{\mu\nu\rho\s}P_\rho W_{*\s}\,.
\label{spinalg}
 \ee
Also as in the VSUSY case, $\hat{W}^\mu $ and $\tilde W_C^\nu $ commute.
However, in this case $\tilde{W}_C^2$ is not a Casimir, in contradistinction to the VSUSY case, where it is proportional to
$P^2$. Despite this fact, it is still possible to use this decomposition in the derivation of the particle content of a multiplet for ordinary supersymmetry.

We have
\begin{equation}
  \tilde W_C^2=-\frac{3}{32}(\bar Q_+Q_+)(\bar Q_-Q_-)+\frac{3}{8}(\bar
Q_+\slP Q_-)\,.
 \label{tilWC2}
\end{equation}
where $Q_\pm$ are chiral projections of the
Majorana super charge
\begin{equation}
  Q_\pm={\cal P}_\pm Q,\qquad {\cal P}_\pm
=\frac12(1\pm i \gamma_5)\,.
 \label{Qpm}
\end{equation}
In terms of these
projections, the odd sector of the ordinary supersymmetry algebra can be rewritten as
follows
\be \left[Q_+^\alpha ,Q_-^\beta \right]_{+}=2({\cal P}_+\slP
C^{-1})^{\alpha \beta }\,,\qquad \left[Q_\pm^\alpha ,Q_\pm^\beta \right]_{+}=0\,.
\ee
The Hilbert space of the theory contains states of three kinds
\be
|Y>,\qquad Q^\A_+ |Y>,\qquad (\bar Q_+Q_+)|Y>\,, \label{hilbert}
\ee
with
\be
Q_-^\A|Y>=0,\qquad \hat W^2|Y>=-P^2\;Y(Y+1) |Y>\,.
\ee
The values of the C-spin for these states are
\bea
\tilde W_C^2|Y>&=&0\,,\nn\\
\tilde W_C^2\,Q^\A_+|Y>&=&\frac{3}{4}P^2\,Q^\A_+|Y>\,,\nn\\
\tilde W_C^2 (\bar Q_+Q_+)|Y>&=&0\,. \label{CspinstatesSUSY}
\eea
Then $|Y>$ and $ (\bar Q_+Q_+)|Y>$ have C-spin 0 and $Q^\A_+|Y>$
have  C-spin $\frac12$.
The Lorentz spin is the sum of Y-spin and
C-spin. In this way, we rederive the well known result that $|Y>$ and $ (\bar
Q_+Q_+)|Y>$ have Lorentz spin $Y$ and
 $Q^\A_+|Y>$
have the Lorentz spins $|Y\pm \frac12|$.
Note the difference with the VSUSY case where the
C-spin is fixed to $ 1/2$ and therefore formula (\ref{spinsum}) allows for a direct
derivation of the particle content of a multiplet. In the ordinary supersymmetry case,
together with the analogous formula (\ref{spindec}), one needs to know also the Hilbert
space structure given by formula (\ref{hilbert}).

\subsection{Odd `Casimir'}

For ordinary supersymmetry, there can be no odd Casimirs, as
the fermions are spinors and hence do not commute with the Lorentz
generators. However, for VSUSY this argument does not hold. We can find
an operator that commutes with all generators of the algebra under one
condition on the mass and central charges, a BPS--like condition. This is
an invariant operator in certain representations of the algebra (we call
it `Casimir' in this paper).

We consider the simplest possibility of an odd operator linear in the
anticommuting generators $(G_\mu,G_5)$. The set of conditions one obtains
when imposing that such operator commutes with $G_\mu$ and $G_5$ admit a
nontrivial solution when the determinant of the following matrix is zero:
\begin{equation}
[G_A,G_B]_+\,=\left[\begin{matrix}{\h_{\mu\nu}Z&-P_\mu\cr-P_\nu&\tilde Z}\end{matrix}\right]\,,\qquad
G_A=(G_\mu,G_5)\,. \label{GGmat}
\end{equation}
The determinant vanishes when the following BPS--like condition is satisfied:
\begin{equation}
Z\tilde Z+m^2=0\,. \label{condZtilZm2}
\end{equation}
In this case, the matrix (\ref{GGmat}) admits the eigenvector $(P_\nu,Z)$
with zero eigenvalue. Therefore, the odd `Casimir' we are looking for has
the form
\begin{equation}
Q= G\cdot P+ G_5Z\,. \label{defQ}
\end{equation}
It commutes with all the generators under the condition
(\ref{condZtilZm2}), implying that when $m\neq 0$ both $Z$ and $\tilde Z$
are nonvanishing and have opposite sign. We can rewrite this condition
in terms of the variable $c$ introduced at the end of section
\ref{ss:VSUSYalg} and the signs of the central charges as
\begin{equation}
c=\sqrt{|Z\tilde Z|}=|m|\,,\qquad\sign(Z)=-\sign(\tilde Z)\,.
\end{equation}
We notice that $Q$ is of course a `Casimir' also when $m=0$, but in that
case one or both central charges must be zero. We are not studying this
case in this paper.

$Q$ acts as an odd constant on the states in the representations
satisfying condition (\ref{condZtilZm2}). Therefore, unless the model
under consideration has a natural odd constant, $Q$ has to annihilate all
states in those representations. As a result, the physical role of the
odd `Casimir' is to give a Dirac-type equation for the particle states.

In principle, one could also look for odd Casimirs that are cubic or of
higher order in the odd generators. In fact, one can prove that such higher
order `Casimirs' do not arise. A brief discussion of this point is
presented at the end of Appendix \ref{app:CasimirsVSUSY}.

\section{VSUSY as a contraction of $\OSp(3,2\,|2)$}
 \label{ss:VSUSYOSp}

It is interesting to explore the connection of VSUSY with other algebras.
Concerning this point, we have already mentioned in the introduction that
the Euclidean version of VSUSY is related to the symmetries of
topological $\mathcal{N}=2$ Yang-Mills theories.

In this section we study the relation of Minkowskian VSUSY with the
simple orthosymplectic algebra $\OSp(3,2\,|2)$. We show that our algebra
arises as a subalgebra of an In{\"o}n{\"u}-Wigner contraction (further we just
write `contraction') of $\OSp(3,2\,|2)$. In fact, the ordinary
supersymmetry algebra can also be derived by a similar contraction
procedure from $\OSp(1|4)$, as shown in \cite{Hlavaty:1980md}.

Both ordinary supersymmetry and VSUSY are generalizations of the Poincar{\'e}
algebra. The Poincar{\'e} algebra itself is not a simple algebra. There are,
however, two well-known connections to simple algebras. First, the
Poincar{\'e} algebra can be seen as a contraction of a simple algebra. The
simple algebra is then the de Sitter algebra, where one introduces a
scale $\lambda $, such that for $\lambda \rightarrow \infty $ the
Poincar{\'e} algebra results using the same generators. Another procedure
starts from the conformal algebra, of which Poincar{\'e} is a subalgebra.

For superalgebras the same facts hold. The ordinary supersymmetry algebra
is a contraction of a super-de Sitter algebra and a subalgebra of a
superconformal algebra, which are both simple superalgebras. Apart from a
few exceptions, the infinite series of superalgebras is either a
generalization of $\U(N)$,
i.e. $\SU(N|M)$ superalgebras, or a generalization of $\SO(N)$, i.e.
$\OSp(N|M)$, which can also be seen as generalizations of $\Sp(M)$.

A tricky point for ordinary supersymmetry is that the fermions should be
in spin representations, while in $\OSp(N|M)$ the fermions are vectors of
$\SO(N)$ and of $\Sp(M)$. Therefore, the bosonic spacetime group in these
superalgebras can not be recognized as the $\SO(N)$ subalgebra of
$\OSp(N|M)$ (with $N$ including both signatures). Instead, we have to use
e.g. for 4 dimensions the equivalences $\Spin(3,2)= \Sp(4)$ for the
(anti)-de Sitter algebra and $\Spin(4,2)=\SU(2,2)$ for the conformal
algebra. Then the superalgebras that can be used are respectively
$\OSp(N|4)$ and $\SU(2,2|N)$  \cite{Nahm:1978tg}, so that the fermions,
being vectors of $\Sp(4)$ or $\SU(2,2)$ are spinors of $\SO(3,2)$ or
$\SO(4,2)$. For VSUSY we do not have this difficulty. We want the
fermions to appear as a vector (or a scalar) of the Lorentz group.
Therefore we will use the embedding in $\OSp(3,2\,|2)$, whose bosonic
part is $\SO(3,2)\times \Sp(2)$.

A brief introduction to $\OSp$ algebras including more technical details
is given in Appendix \ref{app:OSp}. For what follows, it is enough to
know that the $\OSp(3,2\,|2)$ generators are $M_{\m\n}$, $P_\m$, $Z$,
$\tilde{Z}$, $Z'$ (bosonic) and $G_\m$, $S_\m$,  $G_5$, $S_5$
(fermionic). The subset of generators $(M_{\m\n},P_\m, Z, \tilde{Z},
G_\m, G_5)$ has the correct structure to generate the VSUSY algebra.
However, the $\OSp(3,2\,|2)$ commutation relations for the sector of
interest are\footnote{For completeness, the remaining commutation
relations are given in (\ref{remainingcomm}) in Appendix \ref{app:OSp}.}
\begin{eqnarray}
&&\left[M_{\m\n},M_{\rho\s}\right]_-=\eta_{\n\rho}M_{\m\s} +
\eta_{\m\s}M_{\n\rho} - \eta_{\m\rho}M_{\n\s} - \eta_{\n\s}M_{\m\rho}\,,\cr
&&\left[M_{\m\n},P_\rho\right]_-=\eta_{\n\rho}P_\m
-\eta_{\m\rho}P_\n\,,\cr
&&\left[P_\m,P_\n\right]_-=M_{\m\n}\,,\cr
&&\left[M_{\m\n},G_\rho\right]_-=\eta_{\n\rho}G_\m-\eta_{\m\rho}G_\n\,,\cr
&&\left[P_\m,G_\n\right]_-=-\eta_{\m\n}S_5\,,\qquad
\left[P_\m,G_5\right]_-=-S_\m\,,\qquad
\left[G_\m,G_\n\right]_{+}=\eta_{\m\n}Z\,,\nonumber\\
&&\left[G_\m,G_5\right]_{+}=-P_\m\,,\qquad
\left[G_5,G_5\right]_{+}=\tilde{Z}\,,\cr
&&\left[G_\m,\tilde{Z}\right]_-=2S_\m\,,\qquad\left[G_5,Z\right]_-=2
S_5\,,\qquad \left[Z,\tilde{Z}\right]_-=4Z'\,.
 \label{VSUSYnotyet}
\end{eqnarray}
Therefore, it is clear that VSUSY is not a subalgebra of $\OSp(3,2\,|2)$,
but it could arise after a proper contraction.
In order to do this, we rescale the $\OSp(3,2\,|2)$ generators with a
dimensionless parameter $\lambda$ as follows:
\begin{eqnarray}
&&M_{\m\n}\rightarrow M_{\m\n}\,, \qquad Z' \rightarrow Z'\,, \cr &&P_\m
\rightarrow \lambda^2 P_\m\,, \qquad  Z \rightarrow \lambda^2 Z\,, \qquad
\tilde{Z} \rightarrow \lambda^2 \tilde{Z}\,, \qquad \cr && G_\m
\rightarrow \lambda G_\m\,, \qquad G_5 \rightarrow \lambda G_5\,, \qquad
S_\m \rightarrow \lambda S_\m\,, \qquad S_5 \rightarrow \lambda S_5\,,
\label{lambdascale}
\end{eqnarray}
and consider the limit $\lambda\rightarrow\infty$. As a result, the
commutation relations (\ref{VSUSYnotyet}) reduce to the VSUSY algebra
(\ref{vsusyalgebra}).\footnote{We give the contraction
  limit of the remaining commutation relations of $\OSp(3,2\,|2)$ in Appendix \ref{app:OSpcontraction}.}

Therefore, we can conclude that the VSUSY algebra is a subalgebra of the
contraction limit of $\OSp(3,2\,|2)$.

\section{Contraction of the Casimir operators}
 \label{ss:contractionCasimirs}
In this section we derive the Casimir operators of VSUSY by contraction
from $\OSp(3,2\,|2)$.

$\OSp(3,2\,|2)$ has 3 independent Casimir operators \cite{Jarvis:1978bc},
$\CC_2$, $\CC_4$ and $\CC_6$, of the form
\begin{equation}
\CC_n=str(M^n)=\sum (-1)^A {M^A}_B{M^B}_C...{M^C}_D{M^D}_A\,,
\end{equation}
where the generators $M_{AB}$ of $\OSp(3,2\,|2)$ are parametrized as in (\ref{Mparam}) in Appendix \ref{app:OSp}.
Explicitly, the quadratic Casimir reads
\begin{equation}
\CC_2=M_{\m\n}M^{\n\m} + 2P_\m P^\m + 2 \left[G_\m,S^\m\right]+2
\left[G_5,S_5\right]-2Z'^2-\{Z,\tilde{Z}\}\,.
\end{equation}
We do not give explicitly the lengthy formulas for $\CC_4$ and $\CC_6$.
In the following, we have used Mathematica-coding based on the superEDC
package developed by Bonanos \cite{bonanos} to derive and handle such
expressions.

We have shown in the previous section that VSUSY is only a subalgebra of
the contraction limit of $\OSp(3,2\,|2)$. Therefore, when we take the
contraction limit of the Casimir operators of $\OSp(3,2\,|2)$, we obtain
Casimir operators of the contracted algebra, containing VSUSY as a
subalgebra. In order to obtain Casimir operators of VSUSY, we have to
eliminate the extra operators $S_\m$, $S_5$ and $Z'$. For our purposes,
the best way to do this is to introduce a second parameter $\beta$ and
make a rescaling
\begin{equation}\label{beta}
S_\m \rightarrow \beta S_\m\,, \qquad S_5 \rightarrow \beta S_5\,, \qquad
Z' \rightarrow \beta Z'\,.
\end{equation}
By taking the limit $\beta\rightarrow 0$ we reduce to the VSUSY
subsector.

By inspection of the $\lambda$ scaling relations (\ref{lambdascale}), it
is clear that the combinations $P^2$ and $Z\tilde{Z}$ scale with maximal
power in $\lambda$, so that the direct contraction of the three
$\OSp(3,2\,|2)$ Casimirs $\CC_2$, $\CC_4$ and $\CC_6$ can only lead to
combinations of these quantities and explicitly one finds
\begin{equation}
\CC_{n}\rightarrow 2\left( (P^2)^{\frac{n}{2}} - (Z\tilde{Z})^{\frac{n}{2}}\right)\,, \qquad n =
2,4,6 \,,
\label{leadingCn}
\end{equation}
which are clearly Casimirs of VSUSY. Since $Z$ and $\tilde Z$ are central
charges, the new information is that $P^2$ is a Casimir of VSUSY.
Therefore, we can not obtain the VSUSY superspin Casimir $\hat{W}^2$
(\ref{defhW}) from a direct contraction procedure, since $M_{\m\n}$ does
not scale in $\lambda$, see (\ref{lambdascale}), and the corresponding
term in the superspin Casimir does not have maximal order in $\lambda$.

A way out is to start from a homogeneous polynomial in the
$\OSp(3,2\,|2)$ Casimirs $\CC_2$, $\CC_4$ and $\CC_6$ of suitable degree,
characterized by the fact that the terms of maximal order in $\lambda$
exactly cancel out.

Since $\hat{W}^2$ is of order 6 in the VSUSY generators, the simplest
possibility would be to start from a polynomial of order 6. However,
inspection of the general structure of the $\OSp(3,2\,|2)$ Casimirs shows
that a term with two $Z$'s, two $P^\m$'s and two $M_{\m\n}$'s will never
appear, due to the symmetric behavior with respect to $Z$ and
$\tilde{Z}$.

Therefore, we move on to the next nontrivial order by considering a polynomial of
order 8. One can then show that, by imposing the vanishing of the maximal
order terms in $\lambda$ ($\lambda^{16}$), the following combination is
selected uniquely:
\begin{equation}
\mathcal{K}_8=-4\;\CC_6\;\CC_2+3\;\CC_4^2+\frac{1}{4}\;\CC_2^4\,.
 \label{mathcalK8}
\end{equation}
For the contraction limit to give the leading terms of order
$\lambda^{12}$ as a result, it is necessary that not only the
$\lambda^{16}$, but also the intermediate orders $\lambda^{15}$,
$\lambda^{14}$ and $\lambda^{13}$ cancel out. In fact, the terms of odd
power in $\lambda$ are absent, due to the fact that $\mathcal{K}_8$ is
even and only fermionic generators scale with odd powers of $\lambda$. We
have then checked explicitly that the term of order $\lambda^{14}$ can be
reduced to terms of order 12 or lower. To obtain this result, we have
used the (rescaled) commutation relations in (\ref{VSUSYlambdabeta}) -
(\ref{OSpalgebraSSn}), given in Appendix \ref{app:OSpcontraction}, and we
have therefore produced some extra terms of lower order in $\lambda$. The
order 12 terms are the interesting ones for the derivation of the
superspin Casimir because this is the first place where $M_{\m\n}$ terms
appear. We have checked that the extra terms at order 12 generated from
the higher order terms by the use of commutation relations vanish after
the limit $\lambda\to\infty$ is taken. This makes sure that they will not
affect the result of the contraction. We thus have
\begin{equation}
  \mathcal{K}_8= \lambda ^{12}\mathcal{K}_8^{(12)}+ \lambda ^{10}\mathcal{K}_8^{(10)}+\ldots \,.
 \label{CC8splitlambda}
\end{equation}
Up to now we have proven that the contraction limit $\lambda\rightarrow
\infty$ of $\mathcal{K}_8$ gives the order $\lambda^{12}$ term as a
result. We still have to show how this term is connected to the superspin
Casimir $\hat W^2$.

Indeed, one can prove the following relation
\begin{equation}
\mathcal{K}_8^{(12)}=48\left(P^2-Z\tilde{Z}\right)\left(-\frac{\tilde{Z}}{Z}\hat
W^2+\beta^2 \{{\rm terms}~{\rm with}
(S_\m,S_5,Z')\}\right)+f(P^2,Z,\tilde{Z})\,. \label{CC812}
\end{equation}
This analysis proves that $\hat W^2$ is a Casimir operator of the VSUSY
algebra as a consequence of the fact that $\mathcal{K}_8$ is a Casimir of
$\OSp(3,2\,|2)$. Indeed, this includes the statement that any generator
$T$ in the VSUSY algebra commutes with $\mathcal{K}_8$ and hence
\begin{equation}
  0=\lambda ^{-12}\Big[T,\mathcal{K}_8\Big]_-= \left[T,\mathcal{K}_8^{(12)}\right]_- +{\cal O}(\lambda ^{-1})\,.
 \label{proofCasFromOSPstep1}
\end{equation}
We keep in principle the $\lambda $ and $\beta $-dependent commutators in
(\ref{VSUSYlambdabeta}) - (\ref{OSpalgebraSSn}), and the second equality
holds because these do not involve positive powers of $\lambda$. Then we
use (\ref{CC812}) and the fact that $P^2$, $Z$ and $\tilde Z$ were
already recognized as Casimirs of the VSUSY algebra (i.e. they commute
with $T$ up to $\lambda ^{-1}$ terms) and in the leading order of
$\lambda $ commutators with $T$ produce according to (\ref{OSpalgebraGS})
at most terms of order $\beta ^{-1}$. This leads to
\begin{equation}
0= 48\left(P^2-Z\tilde{Z}\right)\left(-\frac{\tilde{Z}}{Z}\left[T, \hat
W^2\right]_-
 +\beta^2b _2+\beta b _1 \right) +{\cal O}(\lambda ^{-1})\,,
 \label{proofCasFromOSP}
\end{equation}
where $b_2$ and $b_1$ are functions of the operators whose explicit form
is not important. Therefore, if we take first the limit
$\lambda\rightarrow \infty$ and then the decoupling limit
$\beta\rightarrow 0$, we obtain that using the commutators of the VSUSY
algebra (i.e. dropping $\lambda ^{-1}$ terms)
\begin{equation}
  \left[ T,\hat W^2\right]_- =0\,.
 \label{ThatW20}
\end{equation}

\section{Conclusions and outlook}
 \label{ss:conclusions}
The aim of this paper is to study the basic algebraic properties of the
VSUSY algebra and its connections with other algebras. Our results will
hopefully shed light on the classification of the irreducible
representations of the algebra, or, at least, will help in the
identification of a class of physically interesting ones. The
representations of VSUSY are not discussed in this paper. We leave this
for future work.

VSUSY shares some common features with ordinary supersymmetry, for
instance the fact that the anticommutator between the fermionic
generators is proportional to the four-momentum $P_{\mu}$. On the other
hand, the fundamental difference between supersymmetry and VSUSY is the
Lorentz nature of their odd generators, spinors for supersymmetry and a
vector and a scalar for VSUSY. We found that in the case $Z,\tilde{Z}\neq
0$ VSUSY has four independent even Casimir operators, $\hat{W}^{2}$,
$P^{2}$, $Z$ and $\tilde{Z}$. We have also been able to construct an odd
operator $Q$, which is nilpotent and behaves like a Casimir when a
BPS--like relation between the central charges and the four-momentum is
satisfied ($Z\tilde Z=P^{2}$).

The Casimir operator $\hat{W}^{2}$ is the square of a Lorentz vector
$\hat{W}_{\mu}$, which is the VSUSY extension of the ordinary
Pauli-Lubanski vector. In the rest frame, it satisfies the $\SU(2)$
algebra and gives rise to the superspin $Y$, the analogue of superspin
for VSUSY. We want to stress that it is necessary to have both central
charges different from zero to ensure that this superspin operator is an
independent Casimir. In fact, in the case $Z=\tilde{Z}=0$ it collapses to
$P^{2}$, up to a constant. On the other hand, the Casimir operator
$P^{2}$ is related to another Lorentz vector, denoted by $W^{\mu}_{C}$.
In the rest frame, $W^{\mu}_{C}$ also satisfies an $\SU(2)$ algebra and
defines a different kind of spin, fixed to the value $\frac{1}{2}$. As a
result of the algebraic relation among the three spin-generating vectors,
a multiplet consists of a doublet of spin $(s,s+1)$ or two spin 1/2
states.

In this paper we have also investigated the relations between VSUSY and
other algebras. First of all, we have observed that an Euclidean version
of VSUSY is a subalgebra of the $\mathcal{N}=2$ topological algebra.
Furthermore, by exploiting the fact that VSUSY displays fermionic
generators which are a vector and a scalar, we have shown how the
(Minkowskian) VSUSY generators are naturally embedded in the simple
orthosymplectic superalgebra $\OSp(3,2\,|2)$.

We have derived the VSUSY algebra and all its independent Casimirs from
$\OSp(3,2\,|2)$ by performing a suitable contraction limit.

The issue of classification of VSUSY irreducible representations remains.
One possibility in this direction is to rewrite the odd sector of the
VSUSY algebra in terms of the generators of a five-dimensional or
six-dimensional Clifford algebra, for which all the irreducible
representations have already been classified. Another possibility would
be to exploit the embedding of the VSUSY algebra in $\OSp(3,2\,|2)$ to
derive the representations. One of the
 final goals would be to construct physically relevant models with underlying VSUSY.
A first example of a particle model is given in \cite{Casalbuoni:2008iy}.
In a field theory context, it would be nice to develop a superspace
formalism for
 VSUSY. In this direction, the connection between VSUSY and $\mathcal{N}=2$ topological
 theories could turn out to be useful, since a (twisted) superspace setup has already
 been constructed
 in that case \cite{Kato:2005fj}. We are currently working on these developments
\cite{workinprogress}.

\section*{Acknowledgments}
We thank Dan Freedman, Gary Gibbons, Satoshi Iso,
Yoshihisa Kitazawa and Tom{\'a}s Ort{\'\i}n for useful discussions. This work has been partially supported by MCYT FPA 2007-66665, CIRIT GC
2005SGR-00564, Spanish Consolider-Ingenio 2010 Programme CPAN
(CSD2007-00042), FWO - Vlaanderen, project G.0235.05 and the Federal
Office for Scientific, Technical and Cultural Affairs through the
`Interuniversity Attraction Poles Programme -- Belgian Science Policy'
P6/11-P.

\appendix

\section{Casimirs  of VSUSY}
 \label{app:CasimirsVSUSY}

In this Appendix we present a general procedure to derive all independent
Casimir operators of VSUSY.

We start from the most general form of an even Casimir operator of the VSUSY
algebra, which reads
\begin{eqnarray}
\CC=C+C^{\mu\nu}G_{\mu}G_{\nu}+C^{\mu 5} G_{\mu}G_{5}
+C^*\ep^{\mu\nu\rho\s}G_{\mu}G_{\nu}G_{\rho}G_{\s}+
C^*_{\mu}\ep^{\mu\nu\rho\s}G_{\nu}G_{\rho}G_{\s}G_{5}\,,
\label{CCgeneral}
\end{eqnarray}
where the coefficients $C$'s are functions of the bosonic generators
$(P,M,Z,\t Z)$ and $C^{\mu\nu}$ is antisymmetric. Any even products of
$G$'s can be arranged  in the above form using the algebra
(\ref{vsusyalgebra}).

The condition $[\CC,G_{5}]_-=0$ implies
\begin{equation}
2C^{\mu\nu}P_\nu-C^{\mu 5}\t Z=0\,,\qquad C^{\mu 5}P_\mu=0\,, \qquad
{4}C^*P_{\mu}+C^*_{\mu}\t Z=0\,,\qquad C^*_{\mu}\ep^{\mu\nu\rho\s}P_\s
=0\,,
\end{equation}
and for non zero $\t Z$ we solve these for $C^{\mu 5}$ and $ C^*_{\mu}$
and obtain
\begin{eqnarray}
\CC&=& C+C^{\mu\nu}\t G_{\mu}\t G_{\nu}+C^*\ep^{\mu\nu\rho\s}\t G_{\mu}\t
G_{\nu} \t G_{\rho} \t G_{\s}\,. \label{CCCgen}
\end{eqnarray}
Here  $ \t G_{\mu}$ is defined by
\begin{equation}
\t G_{\mu}\equiv G_{\mu}+\frac{1}{\t Z}P_\mu G_{5}\,, \label{deftildeG}
\end{equation}
and satisfies
\begin{equation}
[\t G_{\mu},G_5 ]_+=0 \quad {\rm and} \quad [\t G_{\mu},\t G_\nu
]_+=\h_{\mu\nu}Z-\frac{P_\mu P_\nu}{\t Z}\,. \label{commtildeG}
\end{equation}
The invariance of $\CC$ with respect to the Poincar{\'e} subgroup implies
that the $C$'s transform as Lorentz covariant tensors and that they are
functions of $(P_\mu,W_\mu, Z,\tilde Z)$. The covariance requires that
$C$, $C_{\mu\nu}$ and $C^*$ have the following form
\begin{eqnarray}
&&C=C(P^2,W^2,Z,\t Z)\,,\qquad C^*=C^*(P^2,W^2,Z,\t Z)\,,\nonumber\\
&&C^{\mu\nu}=C'(P^2,W^2,Z,\t Z)\ep^{\mu\nu\rho\s}P_\rho W_\s+
C''(P^2,W^2,Z,\t Z)P^{[\mu} W^{\nu]}\,. \label{CCC}
\end{eqnarray}
The remaining condition $[\CC,G_\mu]_-=0$, or, equivalently,
 $[\CC,\t G_\mu]_-=0$ has then to be considered.
We have three Casimirs independent of $\t G$'s,
\begin{equation}
P^2\,,\quad Z\,\quad \mbox{and}\quad  \t Z\,,
\end{equation}
whereas $W^2$ is not a Casimir of the VSUSY algebra.

The most general structure of a Casimir of second order in $\t G$ is
\begin{equation}
\CC_{(2)}=C(P^2,W^2,Z,\t Z) +
C'\ep^{\mu\nu\rho\s}P_\rho W_\s\t G_{\mu}\t G_{\nu} + C''(P^\mu
W^\nu-P^\nu W^\mu)\t G_{\mu}\t G_{\nu}\,. \label{C2G2}
\end{equation}
We start by considering the commutator of the second term with $\tilde
G_\lambda $. After some algebraic manipulations we obtain 
\begin{eqnarray}
[\ep^{\mu\nu\rho\s}P_\rho W_\s \t G_{\mu}\t G_{\nu},\t G_{\lam}]_-=
2Z\left({\ep^{\mu\nu\rho}}_\lam P_\mu W_\nu \t G_{\rho}+
(P^2 \t G_{\lam}-(P\t G)P_\lam)\right) 
=-Z{[}W^2,\t G_\lam]_-\,. \label{C2G213}
\end{eqnarray}
This equation can be written as
\begin{eqnarray}
{[}ZW^2+ \ep^{\mu\nu\rho\s}P_\rho W_\s \t G_{\mu}\t G_{\nu},\t
G_{\lam}]_-&=&0\,.
\end{eqnarray}
Therefore, we have found a Casimir with a second order term in $\t
G_\mu$:
 \begin{eqnarray}
\CC_{(2)}=ZW^2+ \ep^{\mu\nu\rho\s}P_\rho W_\s \t G_{\mu}\t G_{\nu}=ZW^2+
\ep^{\mu\nu\rho\s}P_\rho W_\s G_{\mu} G_{\nu}\,.
\label{C2G4}
\end{eqnarray}
In the previous formula, the $P_\mu$ terms in $\t G_\mu$ do not
contribute. It is convenient to introduce a vector that is a polynomial
in the generators,
\begin{eqnarray}
\hat W^\mu&\equiv& ZW^\mu-\frac{1}{2} \ep^{\mu\nu\rho\s}P_\nu G_{\rho}
G_{\s}=\frac{1}{2} \ep^{\mu\nu\rho\s}P_\nu (ZM_{\rho\s}-G_{\rho} G_{\s})\,.
\label{superspinIIA}
\end{eqnarray}
Its square gives an alternative form of the Casimir operator $\CC_{(2)}$
in (\ref{C2G4}) for the VSUSY algebra since
\begin{eqnarray}
\hat W^2=Z\CC_{(2)} + \frac{3}{4}P^2 Z^2\,.
\end{eqnarray}
The commutator of the $C''$ term in  (\ref{C2G2}) with $\tilde G_\tau$
has a linear and a cubic part in $\tilde G_\mu $. In principle, these
terms could cancel with the contributions coming from the quartic term in
(\ref{CCCgen}). We obtain with a convenient normalization ($C''=1/2$ and
$C^*=\frac{b}{12}\tilde Z $)
\begin{eqnarray}
\lefteqn{\left[P^{[\mu }W^{\nu ]}\tilde G_\mu \tilde G_\nu
+\frac{b}{12}\tilde Z \epsilon^{\mu \nu \rho \sigma } \tilde G_\mu \tilde
G_\nu \tilde G_\rho
\tilde G_\sigma ,\tilde G_\tau \right]_-} \nonumber\\
   &=& ZP^\mu \tilde G_\mu W_\tau -\frac{1}{\tilde Z}\left( Z\tilde
   Z-P^2\right) W^\mu\tilde G_\mu P_\tau \nonumber\\
  &&  +\frac{1}{3}\epsilon^{\mu \nu \rho \sigma }\tilde G_\mu \tilde G_\nu \tilde
   G_\rho \left( (1-b)P_\sigma P_\tau + (b Z\tilde Z-P^2)\eta _{\sigma \tau }\right)
   \,. \label{Casimirtestcombination}
\end{eqnarray}
It is tantalizing that all but the first term cancels for $b=1$ when the
BPS--like condition $Z\tilde Z-P^2=0$ holds. However, the first term
remains and so the hoped cancellation does not occur.

Finally, one may look for higher order odd `Casimirs'.  For $\tilde Z\neq
0$, as we always assume in this paper, in the algebra written in terms of
$\tilde G_{\mu}$ and $G_5$, the only nontrivial commutator involving
$G_5$ is the one between $G_{5}$ and itself. Therefore, requiring that
such a Casimir commutes with $G_5$ implies that $G_5$ can not explicitly
appear, and the only expression that we should look at is
\begin{equation}
  Q^{(3)}= B^\mu \tilde G_\mu + A_\mu \epsilon ^{\mu \nu \rho \sigma }\tilde G_\nu \tilde G_\rho \tilde G_\sigma
  \,,
 \label{Q3parametrized}
\end{equation}
where $A_\mu$ and $B_\mu$ are bosonic vectors, functions of $P_\mu$,
$W_\mu$, $Z$ and $\tilde Z$. We impose that this commutes with $\t
G_\lam$, under the condition (\ref{condZtilZm2}). It can be easily
checked that only $B^\mu =P^\mu $ and $A_\mu=0$ give a solution, which is
the one mentioned in (\ref{defQ}).

In summary, we have shown that there are four  even Casimir operators of
the VSUSY  algebra,
\begin{equation}
  Z\,,\qquad \t Z\,,\qquad P^2\,\quad {\rm and}\quad \hat W^2\,,
 \label{BosonicCasimirs}
\end{equation}
and the odd `Casimir' $Q$ of (\ref{defQ}) in representations satisfying
(\ref{condZtilZm2}),
\begin{equation}
Q= G\cdot P+ G_5Z\,,\qquad P^2-Z\t Z=0\,. \label{defQA}
\end{equation}

\section{Definition and conventions for $\OSp$ algebras}
 \label{app:OSp}

$\OSp$ algebras are very simple generalizations of $\SO$ or $\Sp$
algebras. The $\SO$ algebras have a symmetric metric, the $\Sp$ algebras
have an antisymmetric metric, and the $\OSp$ algebras have a `graded
symmetric' metric.

To understand graded symmetry, one needs the supertranspose of tensors.
If one has graded indices $A,B,...$ which are either bosonic (then
$(-)^A=1$) or fermionic (with $(-)^A=-1$), supertranspose acts
differently according to whether an index is upper or lower. We have
\begin{eqnarray}
 T^{AB} & : & \mbox{supertranspose}:\ (-)^{AB}T^{BA}\,,  \nonumber\\
 T_{AB} & : & \mbox{supertranspose}:\ (-)^{AB+A+B}T_{BA} \,,\nonumber\\
 T^A{}_B& : & \mbox{supertranspose}:\ (-)^{AB+B}T_B{}^A \,,\nonumber\\
T_A{}^B& : & \mbox{supertranspose}:\ (-)^{AB+A}T^B{}_A \,,
 \label{supertranspose}
\end{eqnarray}
i.e., apart from the $(-)^{AB}$ factor, there is an extra factor when a
lower index changes from first to last position. A supertrace is made
with a factor $(-)^A$:
\begin{equation}
  \mathop{str} T = (-)^A T_A{}^A\,,\qquad  \mbox{or}\qquad
\mathop{str} T = (-)^A T^A{}_A\,,
 \label{supertrace}
\end{equation}
which means that this definition is invariant under supertranspose.
Moreover, the supertranspose of the product of matrices $M\,N$ is
$N^T\,M^T$.

A general treatment is given in \cite{DeWitt:1992cy}. The matrices that
we use are all of `bosonic type' in the terminology of this book.

The superalgebra $\OSp$ consists of matrices preserving a graded
symmetric metric $\eta _{AB}$. When we use $\alpha $ for the part of the
indices $A$ that are bosonic and $i$ for those that are fermionic, we can
block-diagonalize such that $\eta _{\alpha i}=\eta _{i\alpha }=0$, $\eta
_{\alpha \beta }=\eta _{\beta \alpha }$ and $\eta _{ij}=-\eta _{ji}$. We
use $\eta_{\alpha\beta}={\rm diag}(-1,+1,+1,+1,-1)$ and $\eta_{21}=1$ for
the $\OSp(3,2\,|2)$ metric. The generators $M_{AB}$ are graded
antisymmetric, which thus means that
\begin{equation}
  M_{AB}=(-)^{(A+1)(B+1)}M_{BA}\,,
 \label{Mgradedantisymm}
\end{equation}
and we can use the inverse of $\eta $ to raise and lower indices, having
the care of putting summed indices always in adjacent positions.

In order to obtain the commutation relations with correct signs we should
use the structures defined above. A convenient way consists in forming a
supertrace of the generators and parameters $\lambda ^{AB}$ which are
graded antisymmetric. This leads to
\begin{equation}
  \left[ M_{AB}, M_{CD}\right] _{\pm}\lambda ^{DC}(-)^C= 2 M_{AD}\lambda ^{DC}\eta
  _{CB} +2(-)^{(A+1)(B+1)}M_{BD}\lambda ^{DC}\eta
  _{CA}\,.
 \label{MMgradedlambda}
\end{equation}
The left-hand side uses anticommutators or commutators, according to the
type of the generators. The right-hand side involves a matrix product of
$M\lambda \eta $ and uses the graded symmetry for the transpose and a
convenient normalization such that the bosonic subalgebra has the usual
normalization for orthogonal algebras.

When we extract the parameters from (\ref{MMgradedlambda}), we have to
respect another graded antisymmetry and  obtain
\begin{eqnarray}
  \left[ M_{AB}, M_{CD}\right] _{\pm}&=&  M_{AD}\eta
  _{CB}(-)^C +(-)^{(A+1)(B+1)+C}M_{BD}\eta
  _{CA}\nonumber\\
  &&+M_{AC}\eta
  _{DB}(-)^{CD+C+1} +(-)^{(AB+A+B+CD+C)}M_{BC}\eta
  _{DA}\,.\label{MMgraded}
\end{eqnarray}
In the special case of $\OSp(3,2\,|2)$ considered in this paper, the generators can be organized in a graded anti-symmetric supermatrix as
\begin{equation}
M_{AB}=\left[\begin{matrix}{ M_{\m\n} & P_\m & G_\m & S_\m \cr -P_\nu  &
0  & S_5 & G_5 \cr G_\nu  & S_5 & Z & Z' \cr S_\nu & G_5 & Z' &-
\tilde{Z}}
\end{matrix}
\right]\,. \label{Mparam}
\end{equation}
In this paper, the $\OSp(3,2\,|2)$ commutation relations rewritten in
terms of the entries of
 this matrix are used.
They are (\ref{VSUSYnotyet}) and the other non-vanishing ones are
\begin{eqnarray}
&&\left[M_{\m\n},S_\rho\right]_-=\eta_{\n\rho}S_\m-\eta_{\m\rho}S_\n\,,\cr
&&\left[P_\m,S_\n\right]_-=-\eta_{\m\n}G_5\,,\cr
&&\left[P_\m,S_5\right]_-=-G_\m\,,\cr
&&\left[S_\m,S_\n\right]_{+}=-\eta_{\m\n}\tilde{Z}\,,\cr
&&\left[G_\m,S_\n\right]_{+}=\eta_{\m\n}Z'-M_{\m\n}\,,\cr
&&\left[S_\m,S_5\right]_{+}=P_\m\,,\cr
&&\left[S_5,S_5\right]_{+}=-Z\,,\cr
&&\left[S_5,G_5\right]_{+}=-Z'\,,\cr
&&\left[G_\m,Z'\right]_-=-G_\m\,,\cr
&&\left[S_\m,Z\right]_-=2 G_\m\,,\qquad\left[S_\m,Z'\right]_-=S_\m\,,\cr
&&\left[S_5,\tilde{Z}\right]_-=2G_5\,,\qquad \left[S_5,Z'\right]_-=-S_5\,,\cr
&&\left[G_5,Z'\right]_-=G_5\,,\cr
&&\left[Z,Z'\right]_-=-2Z\,,\qquad\left[\tilde{Z},Z'\right]_-=2\tilde{Z}\,.
\label{remainingcomm}
\end{eqnarray}

\section{Some useful formulas for the $\OSp(3,2\,|2)$ contraction}
 \label{app:OSpcontraction}
In order to perform the contraction of the $\OSp(3,2\,|2)$ algebra only
the $\lambda$ rescaling (\ref{lambdascale}) is necessary. However, in
order to obtain the Casimirs of VSUSY, as explained in the text, we need
to perform also a $\beta$ rescaling (\ref{beta}). Therefore, we give in
the following the commutation relations rescaled both in
$\lambda$ and $\beta$.
\begin{eqnarray}
&& \left[M_{\m\n},M_{{\rho}\sigma}\right]_-=\eta_{\n{\rho}}M_{\m\s} +
\eta_{\m\s}M_{\n{\rho}} - \eta_{\m{\rho}}M_{\n\s} -
\eta_{\n\s}M_{\m{\rho}}
\,,\nonumber\\
&&\left[M_{\m\n},P_{\rho}\right]_-=\eta_{\n{\rho}}P_\m -\eta_{\m{\rho}}P_\n,\qquad \left[M_{\m\n},G_{\rho}\right]_-=\eta_{\n{\rho}}G_\m-\eta_{\m{\rho}}G_\n\,,\nonumber\\
&&\left[P_\m,P_\n\right]_-=\frac{1}{\lambda^4}M_{\m\n}\,,\qquad
\left[P_\m,G_\n\right]_-=-\frac{\B}{\lambda^2}\eta_{\m\n}S_5\,, \qquad
\left[P_\m,G_5\right]_-=-\frac{\B}{\lambda^2}S_\m\,,\nonumber\\
&& \left[G_\m,G_\n\right]_{+}=\eta_{\m\n}Z\,,\qquad
\left[G_\m,G_5\right]_{+}=-P_\m\,,\qquad
\left[G_5,G_5\right]_{+}=\tilde{Z}\,,\nonumber\\
&& \left[G_\m,\tilde{Z}\right]_-=\frac{\B}{\lambda^2}2S_\m\,,\qquad
\left[G_5,Z\right]_-=\frac{\B}{\lambda^2}2 S_5\,,\qquad
\left[Z,\tilde{Z}\right]_-=\frac{\B}{\lambda^4}4Z'\,.
\label{VSUSYlambdabeta}
\end{eqnarray}
It is then clear that, by taking the limit $\lambda\rightarrow \infty$
with $\beta$ fixed, this sector contracts to VSUSY, given by
(\ref{vsusyalgebra}), which is thus a subalgebra of this $\lambda$-contracted $\OSp(3,2\,|2)$.

The nonzero rescaled commutation relations
between the operators $S_\m$, $S_5$ and $Z'$ and those of the VSUSY algebra are:
\begin{eqnarray}
&&
\left[M_{\m\n},S_{\rho}\right]_-=\eta_{\n{\rho}}S_\m-\eta_{\m{\rho}}S_\n\,, \nn\\
&&\left[P_\m,S_\n\right]_-=-\frac{1}{\B\lambda^2}\eta_{\m\n}G_5\,,
\qquad \left[P_\m,S_5\right]_-=-\frac{1}{\B\lambda^2}G_\m\,,\nonumber\\
&& \left[G_\m,S_\n\right]_{+}=\frac{1}{\lambda^2}\eta_{\m\n}Z'-
\frac{1}{\B\lambda^2}M_{\m\n}\,,\qquad \left[G_\m,Z'\right]_-=-\frac{1}{\B}G_\m\,,\nn\\
&& \left[G_5,S_5\right]_{+}=-\frac{1}{\lambda^2}Z'\,,\qquad
\left[G_5,Z'\right]_-=\frac{1}{\B}G_5\,,\nn\\
&&\left[Z,S_\m\right]_-=-\frac{1}{\B\lambda^2}2 G_\m\,,\qquad
\left[Z,Z'\right]_-=-\frac{1}{\B}2Z\,,\nn\\
&& \left[\tilde{Z},S_5\right]_-=-\frac{1}{\B\lambda^2}2G_5\,,\qquad
\left[\tilde{Z},Z'\right]_-=\frac{1}{\B}2\tilde{Z}\,.
\label{OSpalgebraGS}
\end{eqnarray}
The nonzero commutation relations among these extra generators are
\begin{eqnarray}
 &&
\left[S_\m,S_\n\right]_{+}=-\frac{1}{\B^2}\eta_{\m\n}\tilde{Z}\,,\qquad
\left[S_\m,S_5\right]_{+}=\frac{1}{\B^2}P_\m\,,\qquad
\left[S_5,S_5\right]_{+}=-\frac{1}{\B^2}Z\,,\nn\\
&& \left[S_\m,Z'\right]_-=\frac{1}{\B}S_\m\,,\qquad
\left[S_5,Z'\right]_-=-\frac{1}{\B}S_5\,.
 \label{OSpalgebraSSn}
\end{eqnarray}
The contracted algebra of the $\OSp(3,2\,|2)$ is obtained by taking the
limit $\lambda\rightarrow\infty$ with $\beta=1$. Apart from the VSUSY
algebra (\ref{vsusyalgebra}) the nonvanishing commutation relations are
\begin{eqnarray}
&&\left[M_{\m\n},S_\rho\right]_-=\eta_{\n\rho}S_\m-\eta_{\m\rho}S_\n\,,\cr
&&\left[S_\m,S_\n\right]_{+}=-\eta_{\m\n}\t Z
\,,\qquad\left[S_\m,S_5\right]_{+}=P_\m \,,\qquad
\left[S_5,S_5\right]_{+}=-Z\,,\cr
&&\left[G_\m,Z'\right]_-=-G_\m \,,\qquad
\left[S_\m,Z'\right]_-= S_\m\cr
&&\left[S_5,Z'\right]_-=- S_5 \,,\qquad
\left[G_5,Z'\right]_-=G_5\,,\cr
&&\left[Z,Z'\right]_-=-2Z\,,\qquad
\left[\tilde{Z},Z'\right]_-= 2 \tilde{Z}\,.
\label{commbeta}
\end{eqnarray}

\newpage


\providecommand{\href}[2]{#2}\begingroup\raggedright\endgroup

\end{document}